\documentclass{emulateapj}

\newcommand{\kms}{~km\,s$^{-1}$}
\usepackage[dvips]{color}
\usepackage{amsmath}
\usepackage{ulem}

\shorttitle{A Statistical Study of Prominence Waves}
\shortauthors{Hillier et al.}

\begin{document}
\title{A statistical study of transverse oscillations in a quiescent prominence}

\author{A. Hillier\altaffilmark{1}, R. J. Morton\altaffilmark{2} and R. Erd\'elyi\altaffilmark{3}}

\email{andrew@kwasan.kyoto-u.ac.jp}

\altaffiltext{1}{Kwasan and Hida Observatories, Kyoto University, Kyoto, 607-8471, Japan}

\altaffiltext{2}{Mathematics and Information Science, Northumbria University, Pandon Building, Camden Street, Newcastle upon Tyne, UK, NE1 8ST}

\altaffiltext{3}{Solar Physics and Space Plasma Research Centre (SP$^2$RC), University of Sheffield, Hicks Building, Hounsfield Road, Sheffield, S3 7RH, UK}

\begin{abstract}

The launch of the Hinode satellite has allowed for seeing-free observations at high-resolution and high-cadence making it {well suited} to study the 
dynamics of quiescent prominences. In recent years it has become clear that quiescent prominences support small-amplitude transverse oscillations, 
however, sample sizes are usually too small for general conclusions to be drawn. We remedy this by providing a statistical study of 
transverse oscillations in vertical prominence threads. Over a three-hour period of observations it was possible to measure  the properties of 3436 
waves, finding periods from 50--6000\,s with typical velocity amplitudes 
ranging between 0.2-23~km\,s$^{-1}$. The large number of observed waves allows the determination of the frequency dependence of the wave 
properties and derivation of the velocity power spectrum for the transverse waves. 
For frequencies less than 7\,mHz, the frequency-dependence of the velocity power is consistent with the velocity power spectra generated from observations of the horizontal motions of magnetic elements in the photosphere, suggesting that the prominence transverse waves are driven by photospheric motions.
However, at higher frequencies the two distributions significantly diverge, with relatively more power found at higher frequencies in the prominence oscillations.
These results highlight that waves over a large frequency range are ubiquitous in prominences, and that a significant amount of the wave energy is found at higher frequency.
\end{abstract}


\section{Introduction}

Prominences are cool ($\sim 10000$\,K - \citealp{TH1995}), dense structures ($3$\,-\,$6 \times 10^{11}$~cm$^{-3}$ - \citealp{HIR1986}) that are observed in the corona above the limb.
On disk, prominences are observed as dark absorption structures called filaments.
To support the dense prominence material above the solar surface against gravity, classically it has been the curvature of the magnetic field, creating an upward directed magnetic tension force, that is invoked to counteract the gravitational force \citep{KS1957,KR1974}.

There has been a long history of observation of periodic phenomena in both filaments and prominences \citep[e.g.][]{MOR1960}. 
We focus here on the transverse displacement of the fine-scale threads that make up the prominence, although longitudinal motions are also well-documented (\citealp{VRSetal2007}; \citealp{TRIP2009}). 
Typically, the observed transverse waves are separated into two categories: large-amplitude and small-amplitude oscillations (\citealp{OLI2002, IA2012}). 
The large amplitude oscillations are characterised as being excited by a flare blast wave, generally with velocity amplitudes $>20$\kms, \citep{OKA2004, ISO2007, HER2011, GOS2012}. 
The large-amplitude waves typically show damped sinusoidal motion, similar to that seen in flare-excited kink waves in coronal loops (e.g., \citealp{WHI2012}). 

The small-amplitude waves (velocity amplitudes $\lesssim10$\kms, e.g., \citealp{OKA2007}; \citealp{LIN2007, LIN2009}; \citealp{BER2008}; \citealp{SCH2010}), in contrast, are assumed to be unrelated to flaring activity, although the mechanism(s) that excites these waves still remains a mystery. 
\citet{NING2009} suggested that the observed waves could be driven by the 3- and 5-minute oscillations typically associated with acoustic/slow magneto-hydrodynamic (MHD) perturbations in the photosphere and low corona. 
This suggestion was made as the observed periods were in the range $210-525$~s. 
However, the authors only provide details of 13 periodic waves, so it is hard to draw any conclusions about the excitation mechanism from such a small number of events. 

One mechanism that is frequently suggested for the excitation of transverse waves is the convective motions of photospheric granules. However, the focus is usually directed towards the heating of the solar atmosphere or solar wind acceleration (e.g., \citealp{CRAVAN2005}). 
Wave excitation by convective motions is occasionally named as being a source of the filament/prominence waves (e.g., \citealp{ENG2008}) but, to the best of our knowledge, this idea does not seem to have pursued further. 
It would be a logical way to excite the waves and is pervasive in models for generating transverse waves in the chromosphere/corona \citep[e.g.,][]{CRAVAN2005,MS2010,FED2011,VIG2012}.

In this Letter, we provide the first statistical study of small-amplitude transverse waves in a quiescent prominence. 
The measurements allow for the determination of frequency dependent trends in wave properties and for the derivation of the associated velocity power spectrum for the transverse waves. 
The implications in terms of excitation mechanisms, and the potential role of waves in prominence dynamics are discussed. 

\section{Observations and analysis of thread oscillations}

\subsection{Hinode SOT H-$\alpha$ observations}
The data was obtained by the Hinode Solar Optical Telescope (SOT - \citealp{SUEetal2008}) on the 08 
August 2007 between 18:00:00\,UT and 22:00:00\,UT. 
The target was a quiescent prominence (Figure \ref{prom_slit_fig}) and it was observed by the Hinode 
SOT narrowband imager in H-$\alpha$ line centre ($656.3$\,nm) {with a spectral resolution of $0.009$\,nm}. 
{The observations have a pixel size of $0.16$\,arcsec and a cadence of $10$\,s.} 
The data were processed using standard Hinode SOT 
routines. A frame-to-frame alignment of the time series was performed using cross-correlation, 
providing sub-pixel accuracy on the alignment. Due to a satellite repointing, the data between 
$t=10500$\,s and $11250$\,s of the time series is excluded from further analysis in order to prevent 
the manoeuvre from influencing our results.
{It should be noted that the formation of the H-$\alpha$ line used in this study is still subject for debate.
Recent work on the formation of H-$\alpha$ \citep{LEE2012} implies that for limb observations the intensity is mostly sensitive to the column density, and as shown by \citet{SCH2010} the observed motions are associated with Doppler shifts, meaning that we can work under the assumption that the observed oscillations are as a result of plasma motion. For an in-depth discussion on this topic see \citet{LAB2010}.}

\subsection{Method for determining thread oscillations}
\begin{figure}[ht]
\centering
\includegraphics[scale=0.66, clip=true, viewport=0cm 0cm 15cm 13cm]{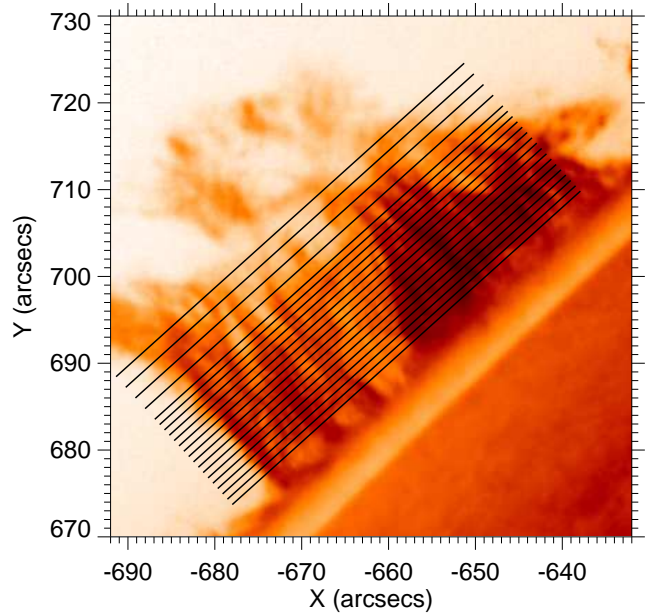}
\caption{A quiescent prominence observed by Hinode SOT on the 8-Aug-2007 at 18:04~UT in H-$\alpha$. The figure shows the negative image of the prominence data. The black lines show the positions of the 21
slits used in the analysis.}
\label{prom_slit_fig}
\end{figure}

\begin{figure*}[ht]
\centering
\includegraphics[scale=0.8, clip=true, viewport=0cm 0cm 22cm 10cm]{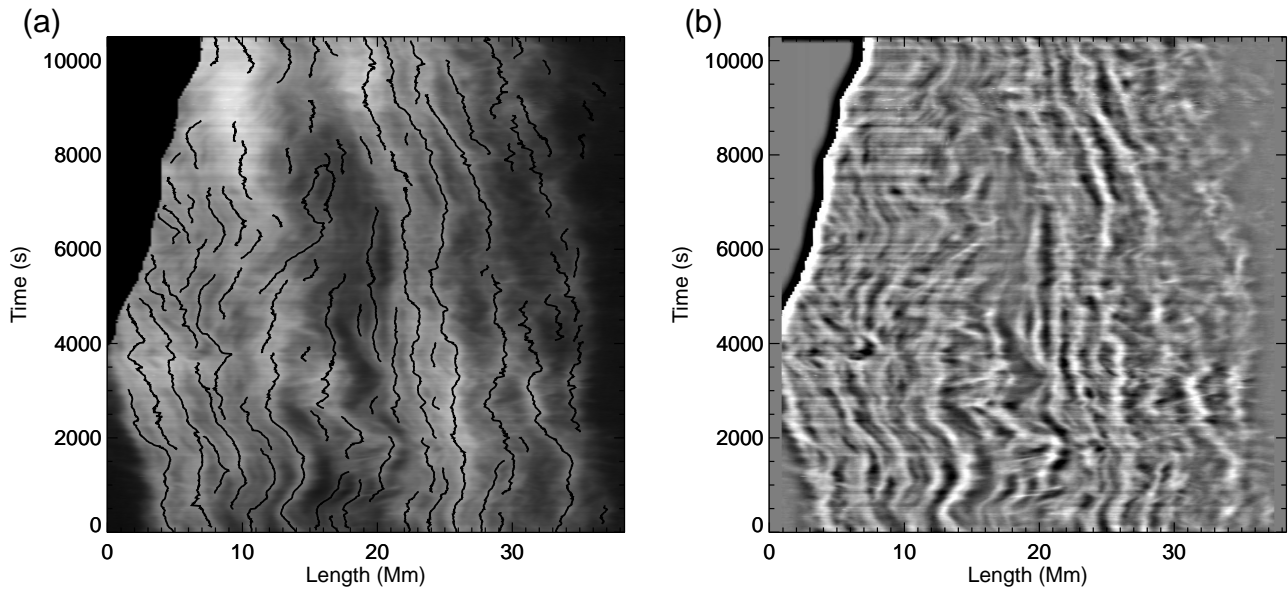}
\caption{Results of the wave tracking routine. Panel a displays the time-distance diagram for slit number 
17 (counting from the bottom slit in Fig.~\ref{prom_slit_fig}) with the over-plotted black lines showing 
the position of the traced threads. Panel b shows an un-sharp masked view of the same slit. The threads 
and their associated transverse motions are clearly evident.}
\label{prom_wave_fig}
\end{figure*}

To observe the transverse displacements of the prominence threads, a total of 21 slits, of
length 54 arcsec, were placed horizontally over the prominence with a separation of 0.83 arcsec, apart from the topmost four slits are spaced at double this separation. To 
increase the position accuracy of the slit pixels, the data pixels were divided into approximately 3 
sub-pixels when constructing the slit. Examples of the slits are given in Fig.~\ref{prom_slit_fig}. Stacking the 
individual slits in time, time-distance diagrams are constructed that show clearly sinusoidal transverse 
displacements of the threads with a range of periods and amplitudes (see, e.g., Fig~\ref{prom_wave_fig}).

To analyse the observed oscillatory motion in the time-distance diagrams, we locate and track the centre 
of the threads using a technique similar to that described in \citet{MOR2013}. For each time-slice of each 
slit, the data is smoothed over 5 sub-pixels to suppress high-frequency intensity fluctuations. For each 
smoothed time-slice of each slit, the sub-pixels with the maximum value of intensity over a range of 30 
sub-pixels (10 pixels) are located. The selected value of intensity has to be the maximum value of 
intensity for a 15-sub-pixel range either side. Any sub-pixel that fulfils this criterion is then tested to 
see if the intensity over the 7 sub-pixels to its left has a positive gradient and the intensity over the 7 
sub-pixels to its right has a negative gradient. If this condition is satisfied, the point is determined as a 
local peak. To improve the accuracy with which the centre of the thread is determined, we take the 
sub-pixel with maximum intensity and the 6 sub-pixels either side, and, fit a quadratic curve to the 
un-smoothed data. The peak of the quadratic curve is then determined to be the central position of the 
thread. We estimate the potential error on the measured position is $0.3$~pixels (private communication 
T. Okamoto).

Next, it is necessary to connect the local peaks, which we assume trace the central position of the vertical 
thread in time. First, a peak in the data is found (the search begins at the first position in the time series) 
and the routine then moves to the next time-step and checks to see if there is a peak contained in the 
seven sub-pixels either side of the peak (and one time-step ahead). This process is repeated for the 
following time-steps until no local peaks are found. If no peak is found, the next five time-steps are 
checked in order. If still no peak is found, the thread is considered to have ended. The selected peaks are 
removed from the data and this process is repeated until all local peaks have been checked for 
connections. In this way, peaks are connected so that their temporal evolution can be followed.
The conditions imposed when connecting peaks would limit any measured velocity amplitudes to 
$<26$~km\,s$^{-1}$. It turns out that this is satisfactory for the present study.

Any time-series of peaks that contains 20 points or more is kept, the rest are discarded. The time-series 
for the peaks of each thread are compared to the time-distance diagrams to remove any false positives. 
Further, the data is checked to see if any two time-series trace the same thread and can be joined 
together. Any gaps in the time-series are filled by linearly interpolating between the data points at the 
beginning and end of the gap and finding the peaks closest to the interpolated line.

An example time-distance diagram is shown in panel (a) of Figure 
\ref{prom_wave_fig}, with the over-plotted black lines showing the traced threads. To highlight the 
threads and their transverse motions, panel (b) shows an unsharped mask version of the data in panel (a).
Running this routine on the 21 slits, the temporal evolution of a total of 3141 sections of the 
vertical threads of the prominence were traced. It should be noted that we use the phrase {\lq sections of 
the vertical threads\rq} because it is not possible to trace all threads for the duration they appear in the 
time-distance plots because of ambiguities that result from the crossing of two threads.

\subsection{Fitting of oscillations}
Before performing a sinusoidal fit to the data, it is necessary to remove any trends. To do this, the data is 
split into its intrinsic mode functions using Empirical Mode Decomposition (EMD) (e.g., 
\citealp{TERetal2004}).
{This method can be viewed as a multi-pass filter that empirically decomposes the time-series data into separate frequency bands.}
The lowest frequency component of an oscillation is subtracted from the 
time-series, removing the trend from the data. The highest frequency fluctuations are also 
removed under the assumption that they are created by noise. This process limits the smallest 
fluctuations that can be fit to periods of $\ge 50$\,s.

\begin{figure*}[ht]
\centering
\includegraphics[scale=0.35, clip=true, viewport=0cm 0cm 23cm 20cm]{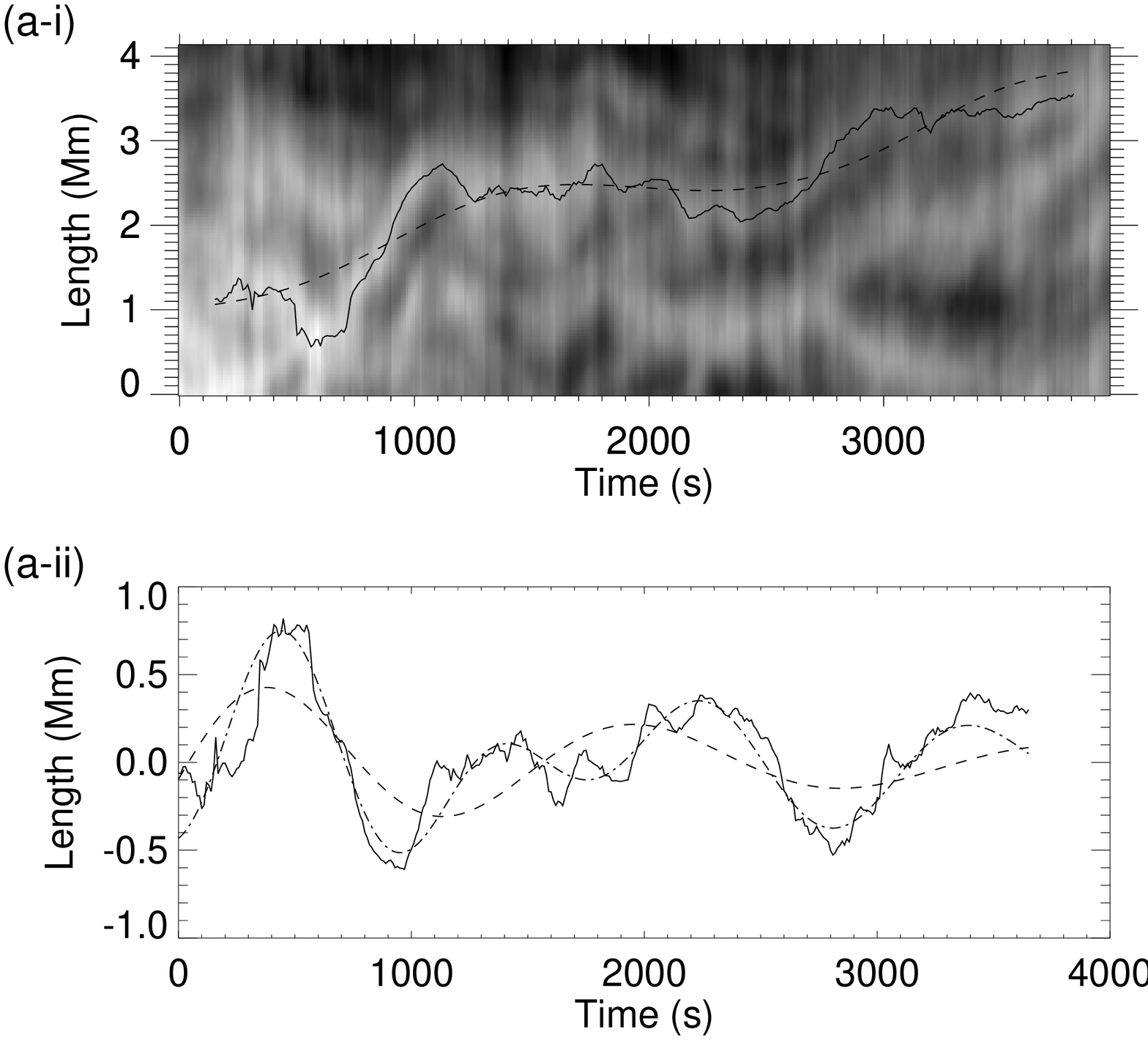}
\hspace{0.1cm}
\includegraphics[scale=0.35, clip=true, viewport=0cm 0cm 23cm 20cm]{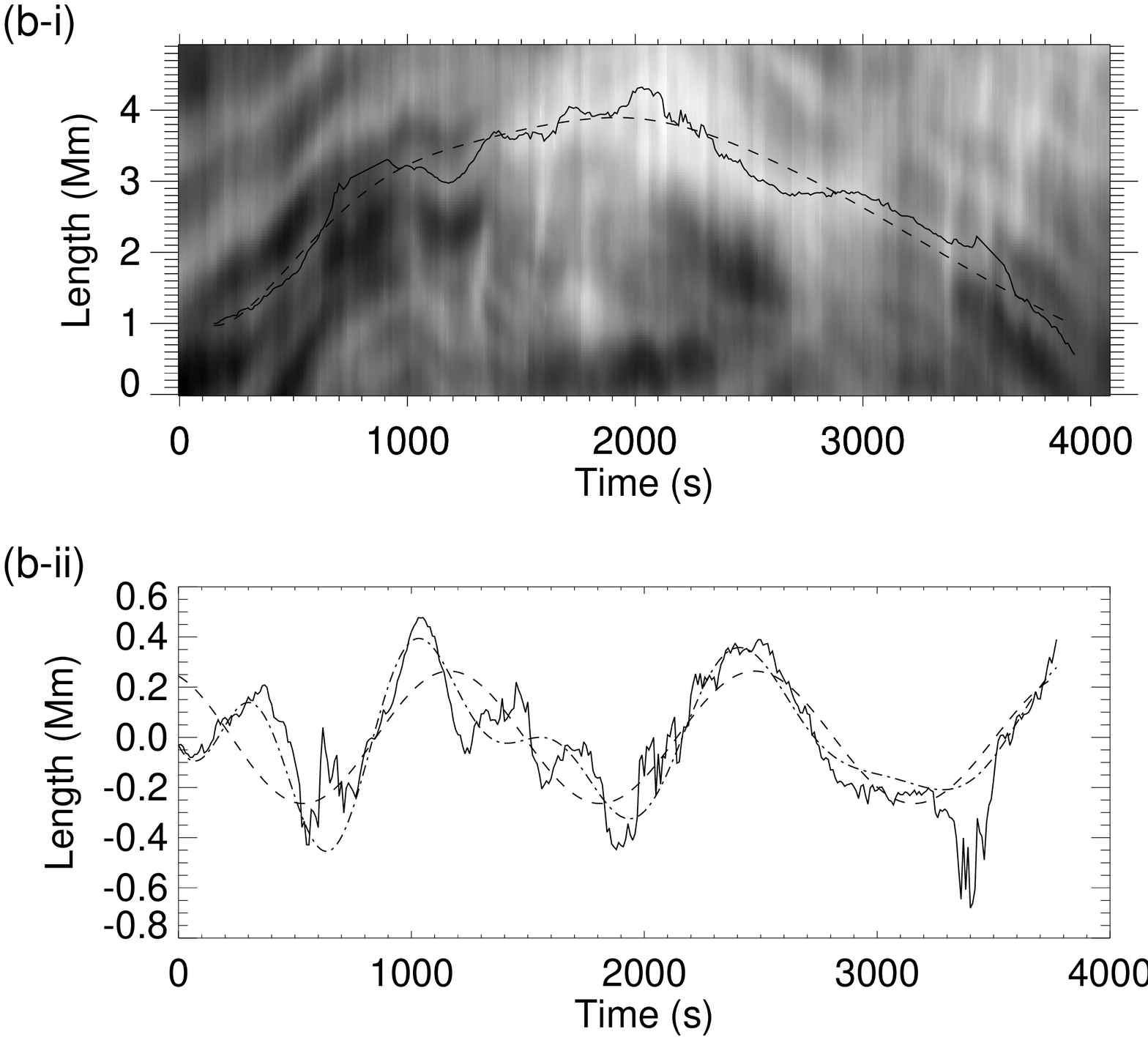}\\
\includegraphics[scale=0.35, clip=true, viewport=0cm 0cm 23cm 20cm]{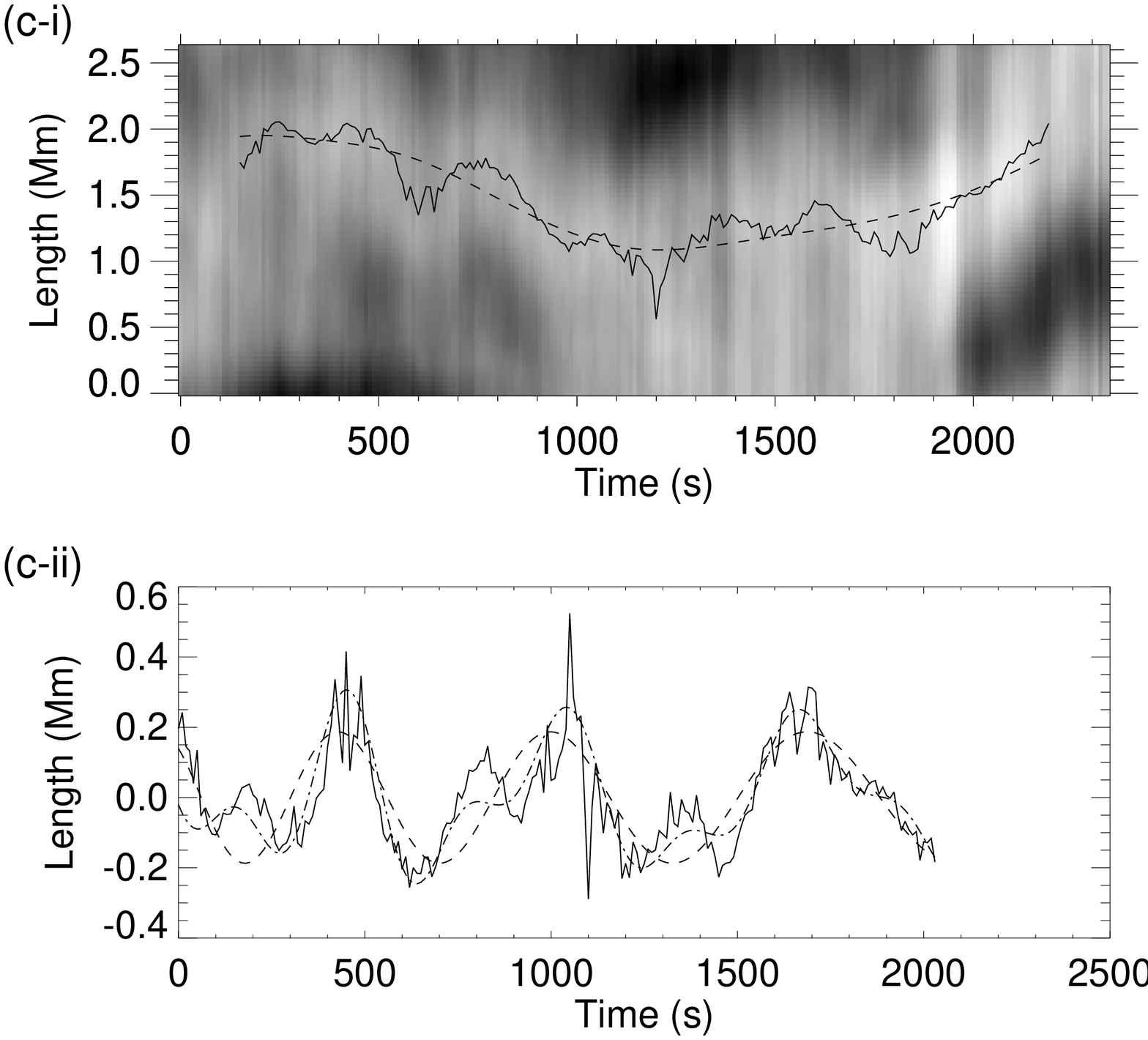}
\hspace{0.1cm}
\includegraphics[scale=0.35, clip=true, viewport=0cm 0cm 23cm 20cm]{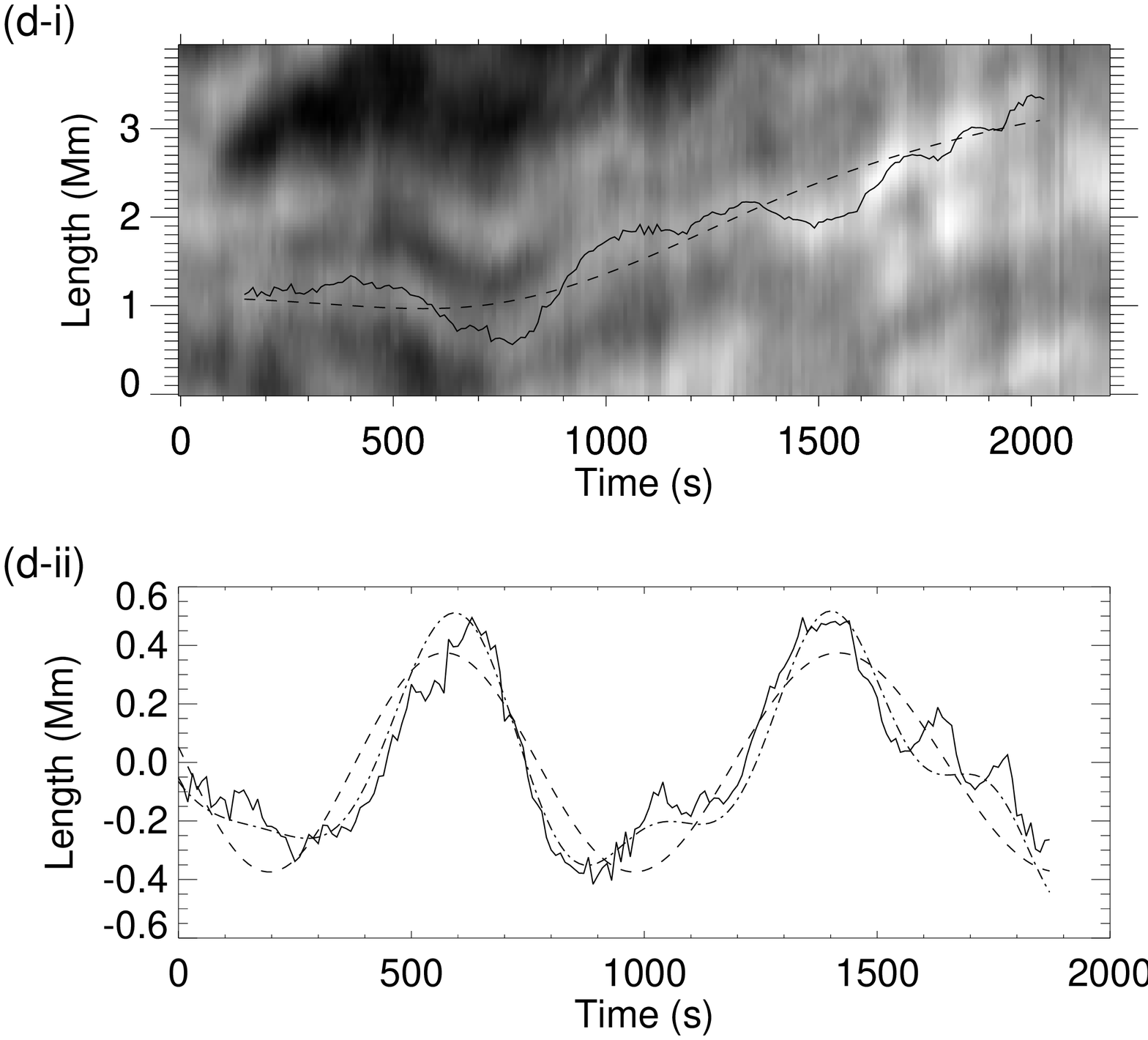}
\caption{The panels a-i to d-i display 4 examples of threads with their centres traced by black lines 
(dashed lines show the trend of each thread found through application of EMD). The panels a-ii to d-ii 
show plots of the detrended curve (solid line), the fit to the initial curve (dashed line) and a combination 
of the original fit and the residual fit (dot-dash line).}
\label{wave_fit}
\end{figure*}

 Any remaining oscillations present are fitted using the following function:
\begin{equation}\label{fitting}
F(t)=A_0 \exp\left( t/\tau \right)\sin\left( \frac{2\pi t}{P_0(1+Ct)} + S \right)
\end{equation}
where $A_0$ is the initial displacement amplitude of the oscillation, $\tau$ is the amplification/damping 
time, $P_0$ is the initial period, $C$ is the linear increase/decrease of the oscillation period and $S$ is 
the phase. This fitting function has been employed by \citet{GOS2012}, 
where the term allowing the change in period with time was found to be of importance in accurately 
determining the period of oscillations. 
Four examples of threads that have been fitted are shown in Figure \ref{wave_fit}. 
The velocity amplitude of the transverse waves is calculated using $V=2\pi A_0/P_0$.

\section{Statistics of transverse prominence oscillations}

\begin{figure*}[ht]
\centering
\includegraphics[width=17cm]{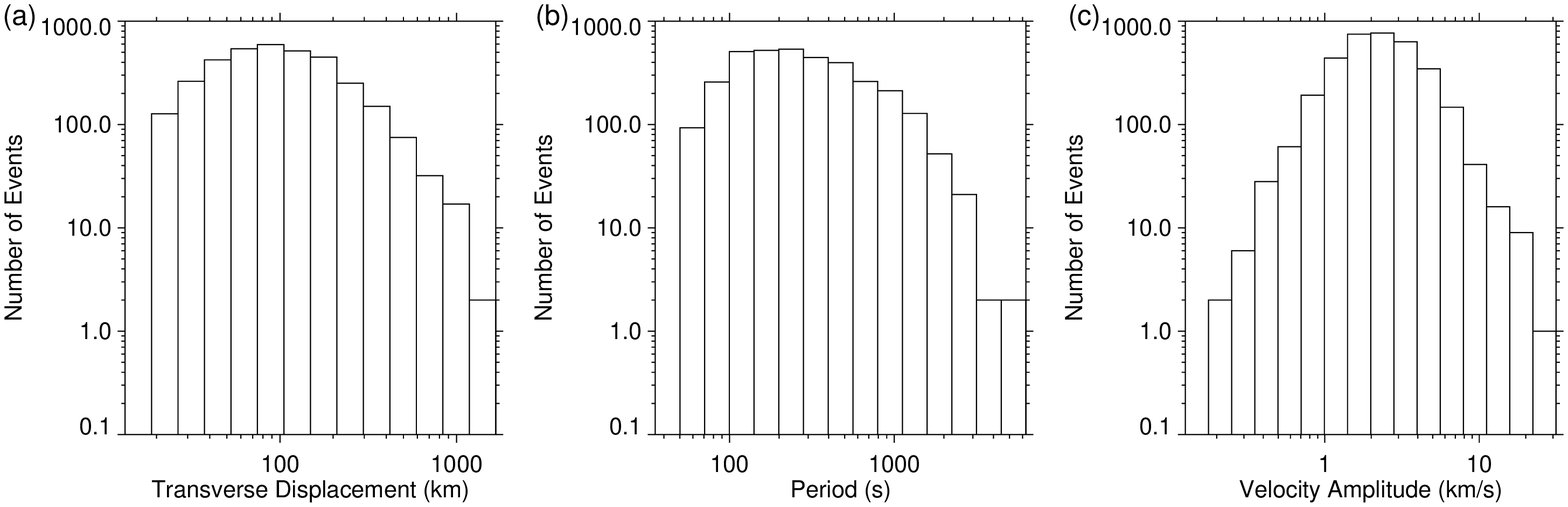}
\caption{Histograms showing the frequency as a function of displacement amplitude ($A_0$), period 
($P_0$) and velocity amplitude ($V$). All histograms are plotted with a log-log scale to allow the ranges 
to be clearly observable. {Only oscillations that display $\ge 1.5$ periods are included in the analysis.}}
\label{wave_histogram}
\end{figure*}

Using the techniques described in the previous section, a total of 3436 oscillations each of more than one and a half periods were 
analysed.
The accumulated statistics for the displacement amplitude, period and velocity amplitude are displayed in 
histograms in Figure \ref{wave_histogram}. The waves have displacement amplitudes 
ranging from 19--1400~km, periods of 50--6000~s and velocity amplitudes of 0.2--23\kms. 
The histograms also reveal that measured events are dominated by the waves with displacement 
amplitudes $<200$~km, periods $<300$~s and velocity amplitudes $<5$\kms.
The histogram for the frequency of events for different velocities shows that the peak frequency occurs at 
approximately V=1-2~\,km\,s$^{-1}$.

\begin{figure*}[ht]
\centering
\includegraphics[width=17cm]{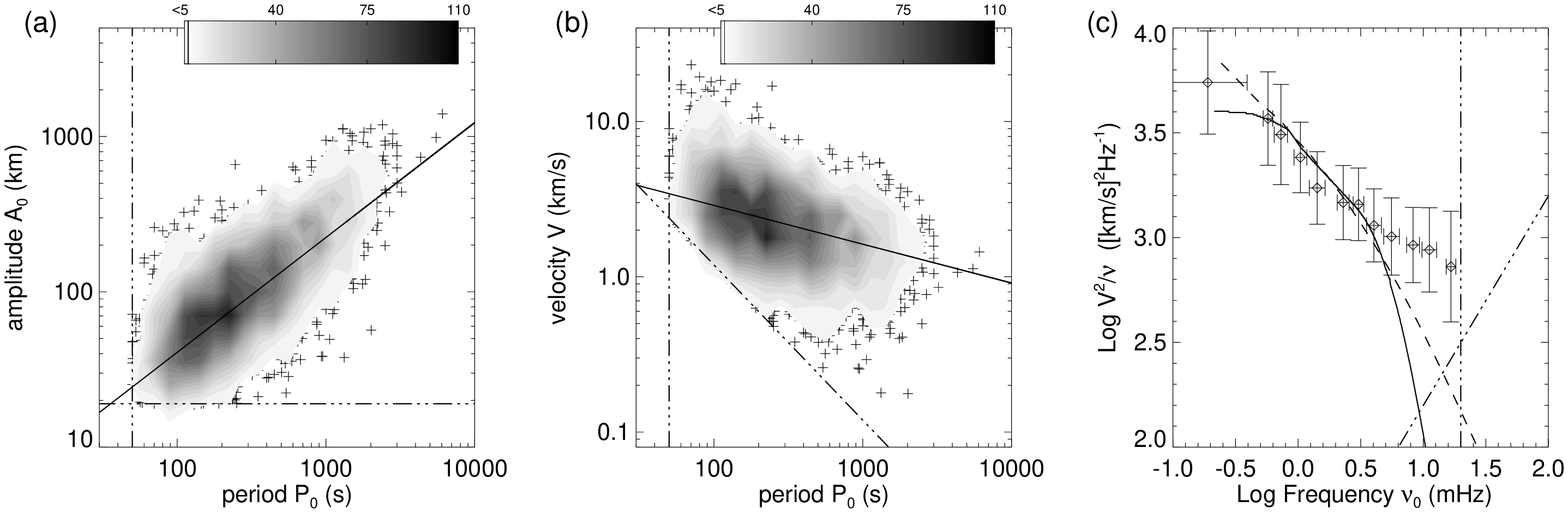}
\caption{Panel (a) displays the displacement amplitude ($A_0$) against initial period ($P_0$), (b) displays 
the velocity amplitude ($V$) against initial period, (c) is velocity power against initial period. The solid 
line in panels (a) and (b) shows the power-law fitted to the data. The solid line and dashed line in panel (c) 
show the velocity power spectrum for horizontal photospheric motions from Figure 3 in \citet{MK2010} 
and the solid line from Figure 9 of \citet{Chitta}. The photospheric power spectra have been scaled for 
comparison with the prominence transverse wave power spectrum. In all panels, the triple-dot-dash line shows the observational limits.}
\label{wave_results_fig}
\end{figure*}

Figure \ref{wave_results_fig} shows plots of (a) the displacement amplitude against the period, (b) the 
velocity amplitude against the period and (c) the  synthesized velocity power spectrum density 
($V(\nu)^2/\nu$) as a function of frequency ($\nu=1/P$). 
{In panels (a) and (b) the contour plot shows the density of the results, with the plus marks showing the positions of the outliers in regions with few points.}
For panels (a) and (b) the straight lines show fitted power laws. 
The power laws found are (a) $A_0=10^{0.13 \pm 0.02}P_0^{0.74 \pm 0.04}$ and (b) $V=10^{0.96 \pm 0.02}P_0^{-0.25 \pm 0.04}$. 
The observations and the analysis technique come with constraints on the minimum period (50~s) and minimum displacement amplitude 
(19~km) we are able to measure. These constraints also affect the minimum measurable velocity 
amplitude. The respective constraints are over-plotted in each plot (dash triple dot). 

The data points plotted in Figure \ref{wave_results_fig}c were obtained by binning the data into 
log-frequency bins of width $\log\nu\sim0.15$. Any bin with less than 30 events was amalgamated with a neighbouring bin. The data for each frequency 
bin was binned into log-power spectrum bins, the size of the bin was varied to some degree due to the 
different number of events in each frequency bin but was generally kept to be approximately 0.15. 
The histogram of the binned power spectrum density for each frequency bin was then fitted with a 
Gaussian distribution, with the central peak of the Gaussian distribution taken as the centre of the 
probability density function of the data.
The errors bars show two times the standard error of the fit (i.e., $\sigma/\sqrt{N}$ where $\sigma$ is the standard deviation of the results and N is the number of bins). 
The position of the data points along the frequency axis and the range (horizontal bars) were determined 
by the mean and the standard deviation of the data points in the frequency bin. 
The Gaussian distribution provides a very good representation of the distribution 
of the power spectrum density in each frequency bins, implying that the data is log-normally distributed. 
The triple dot-dash lines show the observational limits of the study.

Perhaps most interestingly, in Figure \ref{wave_results_fig}c, we over-plot the measured photospheric 
power spectra derived from granulation (solid line - \citealp{MK2010} - multiplied by a factor of 
$10^{1.1}$) and from magnetic bright point motions (dashed line - \citealp{Chitta} - multiplied by a 
factor of $10^{1.4}$), which have been shown to correspond to the motion of concentrations of magnetic flux both observationally \citep[e.g.][]{BERG96} and numerically \citep[e.g.][]{SCHUS03, SHEL04}.
Note, the photospheric velocity power is scaled upward to allow for a direct comparison. The difference in magnitude of velocity power between the photosphere and the 
prominence is a reflection on the increase in velocity amplitudes of waves with height, namely due to 
lower values of magnetic field strength and density in the prominence than in the photosphere, i.e., the 
increase in Alfv\'en speed.

The distribution of the synthesized power spectrum for the prominence waves and the observed 
photospheric power spectra show a number of similarities and differences. The power distribution for 
low-frequency waves can be seen to closely follow the photospheric power spectra, 
although, for higher frequencies (greater than $7$\,mHz), the distributions diverge. 

\section{Implications for Prominence Energetics}

We present here the largest study to date of transverse waves supported by vertical quiescent prominence threads in a single prominence, measuring 3436 periodic motions, which allows us to draw general conclusions about the properties of transverse waves in prominences. 
Firstly, we can confidently state that both the displacement amplitude and velocity amplitude are functions of period. 
The variation in displacement amplitude and velocity amplitude with period appear to qualitatively agree with results found from large-scale studies of transverse waves in chromospheric fibrils (\citealp{MOR2013,MORetal2013b}). 

Secondly, and perhaps most interestingly, we can hypothesise about the excitation mechanism of the prominence waves. 
The horizontal motions of the photosphere has long been conjectured to be the driver of incompressible waves including prominence waves, however, no direct evidence of this has been provided by observations. 
In Figure~\ref{wave_results_fig}c we compare the velocity power spectrum of the observed transverse oscillations with the velocity power spectrum of horizontal photospheric granular motions and horizontal magnetic bright point motions. 
There are also slight differences between the power spectra derived in \citet{MK2010} and \citet{Chitta}, e.g., magnitudes of the velocity power, but the profile is very similar. After scaling the photospheric distributions, it is clear that that the velocity power spectrum of the prominence waves correlates well with the photospheric power spectra for $f<7$~mHz. 
We suggest this result provides evidence for the horizontal motions being the driver of prominence waves. However, further information (e.g., study of phase spectra) is required to attribute full
causality to the horizontal motions. 
There is no evidence for enhanced power at 3 or 5 minutes, potentially ruling out {\it p}-modes as the main driving mechanism.

One further interesting point of note is that the prominence waves show greater velocity power in the high-frequency range than the photospheric spectra. 
A similar result was found for the analysis of fibrils in the solar chromosphere by \citet{MORetal2013b}, implying that the position of the break in the power law \citep[found to be approximately $5$\,mHz in][] {MK2010} could be at higher frequencies in many cases or that the chromosphere processes the energy in such a way that the higher frequency component is enhanced relative to the lower frequency component. 
Either way, the existence of the relatively enhanced power at high frequencies should be of great interest for studying the transport of energy by waves from the photosphere to the corona. 
We note that the results in this paper cannot rule out that local instabilities in a prominence, e.g. the convective continuum instability \citep{BK2011}, driving some of the observed prominence oscillations.

The prospect that the photospheric motions are driving the observed prominence waves raises a number of questions regarding wave propagation through the lower solar atmosphere. 
What causes the deviation from the photospheric power spectra at higher frequencies, be it an enhancement at higher frequencies or a sink reducing the power at lower frequencies, and why the power of the velocity power spectra for frequencies $<7$~mHz remains unaffected after propagation through the chromosphere both need to be understood.
Studies show that wave spectra can be modified due to mode conversion (e.g., \citealp{CARBOG2006}; \citealp{FED2011}) at altitudes where the plasma $\beta$ equals unity or due to reflection from strong gradients in plasma quantities present in the transition region (e.g., \citealp{CRAVAN2005}; \citealp{FED2011}). However, the physical processes involved are complex, meaning simulations of wave propagation from the photosphere to the prominence are likely required to answer such questions.

Finally, we pick up on a comment by \cite{VB2010} related to prominences, namely, {\lq Others have 
suggested that... the plasma is supported by MHD waves. However, relatively high frequencies and wave 
amplitudes are required, and it is unclear why such waves would not lead to strong heating of the 
plasma.\rq} 
Here we demonstrate that the prominence plasma does indeed support numerous MHD waves over a wide frequency range, with velocity power comparable to those seen in chromospheric fine structure (\citealp{MOR2012b}), who suggest that the chromospheric waves have an energy flux capable of maintaining a heated atmosphere. 
The following questions then naturally arises: what happens to the wave energy observed in prominence fine-structure? 
If the MHD waves are suitable for the heating of the solar atmosphere, why does the prominence plasma not display evidence for strong heating? At present we cannot answer these questions but they are clearly a future challenge for that models of prominences should aim to answer.

\acknowledgements{The Authors would like to thank the anonymous referee whose comments improved the manuscript. AH is supported by KAKENHI Grant-in-Aid for Young Scientists (B) 25800108. RM is grateful to the Northumbria University for the award of the Anniversary Fellowship and the Royal 
Astronomical Society for Travel Grants. RE acknowledges M. K\'eray for patient encouragement and is also grateful to NSF, Hungary (OTKA, Ref. No. K83133). This work is supported by the UK Science and Technology Facilities Council (STFC).
The authors would like to thank L. P. Chitta and T. Matsumoto who provided us with the photospheric power spectra results and K. Otsuji for providing the image alignment routines.
Also, thanks is given to J-L. Ballester and I. Arregui for their constructive comments that greatly improved the quality of this manuscript.
AH acknowledge the support by the International Space Science Institute (ISSI, Switzerland) and discussions within the ISSI Team 214 on Flow-Driven Instabilities of the Sun–Earth System.
Hinode is a Japanese mission developed and launched by ISAS/JAXA, with NAOJ as domestic partner and NASA and STFC (UK) as international partners.
It is operated by these agencies in co-operation with ESA and the NSC (Norway)
}

\end{document}